\pdfoutput=1
\documentclass[a4paper]{article}
\usepackage{blindtext}
\usepackage{cite}
\usepackage{authblk}
\usepackage{multicol}
\usepackage[margin=0.8in]{geometry}
\usepackage{graphicx}
\usepackage{float}
\usepackage{titlesec}
\usepackage{multirow}
\usepackage{caption}
\usepackage{hyperref}

\titleformat{\section}
{\Large\bfseries}
{}
{0em}
{\hrule height 1pt \vspace{0.5em}\filcenter}
[\vspace{0.5em}\hrule height 1pt]

\title{\textbf{AI-based Self-healing Solutions Applied to Cellular Networks: An Overview}}
\author{\textbf{Jaleh Farmani} \thanks{Sapienza University of Rome} \and \textbf{Amirreza Khalil Zadeh}\thanks{Tor Vergata University of Rome}}

\affil{

\texttt{farmani.1989067@studenti.uniroma1.it \\
\texttt{amirreza.khalilzadeh@students.uniroma2.eu}
}}
\date{February 2023}
\begin{document}

\maketitle

\begin{multicols}{2}

\textbf{\textit{Abstract} — In this article, we provide an overview of machine learning (ML) methods, both classical and deep variants, that are used to implement self-healing for cell outages in cellular networks. Self-healing is a promising approach to network management, which aims to detect and compensate for cell outages in an autonomous way. This technology aims to decrease the expenses associated with the installation and maintenance of existing 4G and 5G, i.e. and emerging 6G networks by simplifying operational tasks through its ability to heal itself. We provide an overview of the basic concepts and taxonomy for SON, self-healing,  and ML techniques, in network management. Moreover, we review the state-of-the-art in literature  for cell outages, with a particular emphasis on ML-based approaches.
}

\section{\Roman{section}. Introduction}
 
The evolution of mobile networks from one generation to another has been mainly driven by hardware technology advancements. However, the transition to 6G is different as novel software technology advancements will be crucial, particularly in the network management area. As the standardization process 6G is still in the early stages, so the industry will likely continue to look for cost-effective solutions that will provide the same level of performance as that provided by state-of-the-art architectures. There are several research initiatives aimed at addressing this challenge. Some of the proposed solutions rely on techniques such as machine learning and deep learning, which are being successfully used in many different applications, including autonomous vehicles, image processing, and robotics\cite{janiesch2021machine}. 

In the case of cellular networks, the main goal is to enable the network to make decisions and automatically handle certain issues without the need for human intervention. With the use of machine learning and autonomous network management, the concept of self-awareness, self-configuration, self-optimization, and self-healing can be achieved. 

The focus of this article is on the self-healing concept and we start by the description of Self-organizing Network (SON). The main objective of SON is to bring intelligence and autonomous adaptability into cellular networks, reduce operational (OPEX) and capital expenditures (CAPEX) and enhance network performance\cite{moysen2017self}. 
 
 This paper is organised as follows. First, in section II, we describe the concept of SON and how it can be used to optimize the configuration and management of cellular networks. In section III, we provide an overview of the concept of self-healing in cellular networks, including cell detection and compensation. 
Section IV discusses the various machine learning techniques that are commonly used in self-healing, including supervised, unsupervised and reinforcement learning and also some variations of deep learning methods. In section V, we introduce specific AI-based self-healing solutions that have been successfully applied in cellular networks and will briefly describe the methods used. Finally, Section VI concludes the survey.

\section{\Roman{section}. Self-Organizing Networks (SON)}

The concept of SON or Self-Organizing Networks is focused on enhancing the overall performance of cellular networks by bringing in intelligence and adaptability to the system while reducing human intervention. This is done to improve network capacity, coverage and service quality while also simplifying operational tasks such as configuration, optimization and troubleshooting. The main motivation behind implementing SON is two-fold. Firstly, the market demands a diverse range of services and there is a need to reduce time to market of innovative services, leading to increased competition and the need for cost reduction. Secondly, the technical complexity of future radio access technologies imposes significant operational challenges due to the multitude of tunable parameters and dependencies among them. The advent of heterogeneous networks is expected to increase the number of nodes in this new ecosystem, making traditional manual and field trial design approaches impractical.
Although many SON solutions are available, there are still open challenges that need to be addressed, such as the problem of coordination of SON functions and the proper solution of the trade-off between centralized and distributed SON implementations (H-SON)\cite{moysen20184g}. Additionally, the management of 6G networks is expected to provide a new set of challenges due to the need to manage future network complexity, support increased traffic and users, and improve energy efficiency and user experience. The article\cite{chaoub2021self} proposes that SONs and H-SONs can play a crucial role in addressing these challenges and achieving efficient network management and control in 6G networks. Despite the growing interest in 6G, it is still in its early stages, and researchers have only just begun to develop its design, explore its implementation, and consider potential use cases. As a result, the focus has shifted from theoretical AI studies to practical deployment and standardization, as both academic and industrial communities work to move beyond conceptualization. However, there is a lack of research on creating an end-to-end framework for AI distribution that would make it easier to access services through third-party applications, while also utilizing zero touch service provisioning. \cite{baccour2023zero}
\begin{figure}[H]
    \centering
    \includegraphics[width=6cm]{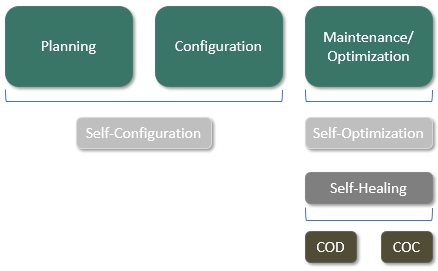}
    \captionsetup{font=small}
    \caption{Self-Organizing Networks}
    \label{fig:1}
\end{figure}

The concept of SON in mobile networks can also be divided into three main categories. As depicted in Figure 1, These categories are: self-configuration, self-optimization and self-healing and are commonly denoted jointly as self-x functions\cite{aliu2012survey}. The focus of this article is the self-healing category.

\begin{figure}[H]
    \centering
    \includegraphics[width=8cm]{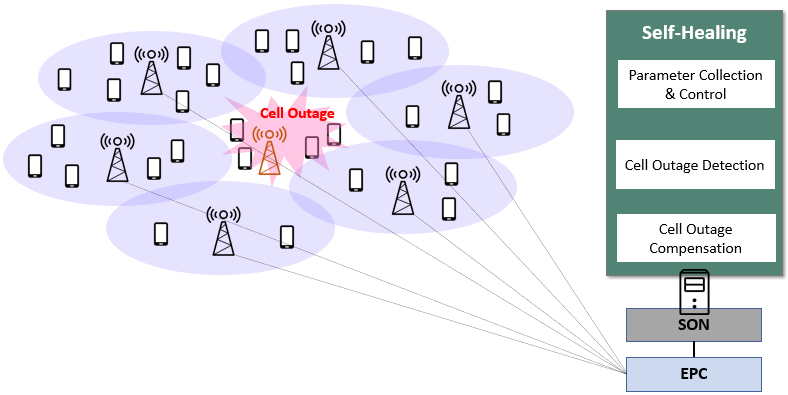}
    \captionsetup{font=small}
    \caption{Overview of the COM framework. The center
site is in outage.}
    \label{fig:2}
\end{figure}
\section{\Roman{section}. Self-healing}
Self-healing technology aims to maintain the wireless cellular networks during the maintenance phase, and it becomes more crucial in 5G and 6G networks due to their highly complex and heterogeneous nature. These networks are also prone to faults and failures, and the most critical domain for fault management is the Radio Access Network (RAN). Each network element serves a specific area, making redundancy a challenging task. If a network element fails to fulfill its responsibilities, it results in degraded performance, and users suffer from poor service quality, resulting in a considerable revenue loss for the operator. To address this challenge, self-healing technology has been introduced in emerging cellular networks. It focuses on the automatic detection of network faults and failures and the implementation of the required corrective actions to mitigate the service degradation effect.  The main defined use cases are the following.

\textbf{\textit{1) Cell Outage Management (COM)}} This use case is split in two main functions:

    \textbf{\textit{1.1) Cell Outage Detection (COD):}} This function automatically identifies cell outages by using input parameters such as Key Performance Indicators (KPIs), alarms, and measurements. When the values of these parameters meet the COD condition, the cell outage is detected. For example, if the value of a KPI exceeds a predefined threshold or an alarm is triggered during a cell outage, COD function identifies the outage.

    \textit{\textbf{1.2) Cell Outage Compensation (COC):}} This  function automatically compensates for a cell outage to maintain the cell operations. When a neighboring cell detects a fault, it classifies the type of fault and makes a compensation decision. The compensation can be in the form of relay-assisted handover (HO), power compensation, or reconfiguration of their antenna tilt. These compensation techniques help to mitigate the impact of the outage on the network and maintain the quality of service for the users.

\section{\Roman{section}. ML techniques used in COM}
This section outlines the different categories of problems that need to be addressed for self-healing of wireless cellular networks. For each category of problem, we identify the machine learning techniques that can be employed. The primary objective of using machine learning is to enhance the performance of specific tasks by creating a model that can identify patterns through learning algorithms. Machine learning taxonomy traditionally comprises three categories: 1.Supervised Learning (SL), 2.Unsupervised Learning (UL), and 3.Reinforcement Learning (RL). Recent advancements in software engineering, computational capabilities, and memory availability have led to the emergence of new trends in the field of machine learning, with deep learning being proven to be effective in a wide range of applications, including language, video, speech recognition, and object and audio detection, among others.
The classes of problems that need to be addressed when managing the network autonomously are:
\begin{itemize}
    \item Anomaly detection: This is a method commonly used in the diagnosis of network faults or misbehaviors, as it is designed to identify and flag abnormal behavior within the network. This approach relies on unsupervised learning or supervised learning methods and is particularly useful when it comes to detecting issues related to faults or improper network settings. Popular algorithms for anomaly detection include k-NN, Local Outlier Factor\cite{amirijoo2009cell} and DBSCAN algorithm that is used in \cite{ali2018self}.
\item Pattern identification: Pattern identification is a task that involves identifying patterns or groupings of cells that exhibit similar behavior, such as increased traffic or dropped calls. This kind
of problems relates to COD issues and solutions
can be found in UL and SL literature. For instance, Decision Tree is a technique that can be used to classify cells into different groups based on their characteristics, such as traffic load and signal strength\cite{omar2021novel}.
\item Control optimization: This category of problems arises frequently in the realm of autonomous network management, where control decision problems are encountered to adjust network parameters online, with the goal of achieving specific performance targets. These types of decision problems, where the optimal decision is learned online based on the feedback from the environment to the network's actions, can be tackled using RL\cite{onireti2015cell}, \cite{moysen2014reinforcement}, and DRL techniques\cite{guo2019deep}, \cite{yu2020deep}. COC problems can be solved using these methods.

\end{itemize}
The upcoming section will provide selected ML techniques which have been used in cell outage management studies or have shown potential for cell fault management from their successful application in similar work. It should be noted that the list of machine learning methods presented is not complete, and certain techniques may not be mentioned. The purpose of this taxonomy is to provide a helpful framework for approaching self-healing problems.

\textbf{A. Supervised Learning (SL): }

The supervised learning (SL) ML technique is useful when the network management (NM) function requires estimation, prediction, or classification of variables. SL is a type of machine learning that can be used for both classification and regression tasks. In classification, the goal is to predict a discrete label or category based on the input features. The output variable is categorical, and the algorithms used for classification include logistic regression, decision trees and support vector machines. In regression, the goal is to predict a continuous value based on the input features. The output variable is continuous. As an example linear regression can be mentioned. In both classification and regression, the model is trained using labeled data to learn a mapping function $f:X\rightarrow Y$ that maps input data $X$ to output labels $Y$ with high accuracy.
In the context of self-healing, SL
algorithms are applied mainly on fault detection use cases.
In the following we
briefly introduce the most common algorithms.

\textbf{1) k-Nearest Neighbors (k-NN)} is a simple but effective classification and regression algorithm. The algorithm works by finding the \textit{k} closest data points in the training set to the new input and classifying or predicting the label of the input based on the labels of its nearest neighbors.

k-NN is a non-parametric algorithm, meaning that it does not make assumptions about the underlying distribution of the data. KNN is also flexible as it can work with any distance metric, such as Euclidean distance or cosine similarity, to measure the similarity between data points. However, KNN can be computationally expensive for large data sets, as it requires calculating the distance between the new input and all training data points. KNN is also sensitive to the choice of the number of neighbors, K, which can affect the bias-variance trade-off of the algorithm. In the
case of cellular networks, k-NN is generally applied in the
context of self-healing, by detecting outage \cite{onireti2015cell}.

\textbf{2) Naive Bayesian}. Naive Bayes is a classification algorithm based on Bayes' theorem, which calculates the probability of a hypothesis given evidence. Naive Bayes assumes that the input features are independent of each other, hence the term "naive". Despite this simplifying assumption, Naive Bayes has been shown to be effective in many real-world applications, such as text classification and spam filtering. Naive Bayes works by first learning the probabilities of each class based on the input features. Then, for a new input, it calculates the probability of each class and chooses the class with the highest probability as the predicted label. Naive Bayes is a fast and efficient algorithm that can handle large data sets and high-dimensional feature spaces. Recent
research has applied the concept of Bayes’ classifiers in fault detection \cite{fortes2015contextualized}.

\textbf{4) Support Vector Machines (SVMs)}is a popular machine learning algorithm for classification and regression tasks. In SVM, the goal is to find a hyperplane that separates the data points into different classes with maximum margin, i.e., the distance between the hyperplane and the nearest data points of each class is maximized. The data points closest to the hyperplane are called support vectors, and they are used to define the hyperplane. SVM can handle both linearly separable and non-linearly separable data by using a kernel trick to map the input features into a higher-dimensional space where the data points become separable. SVM has several advantages, including good performance on high-dimensional data, ability to handle non-linear decision boundaries, and robustness to outliers. However, SVM can be sensitive to the choice of hyper-parameters, such as the regularization parameter and kernel function, and can be computationally expensive for large data sets.
(SVMs) are being used in self-healing scenarios in cellular networks, particularly for fault detection \cite{ciocarlie2013detecting},\cite{zoha2015data} and \cite{zoha2016learning}.

\textbf{5) Decision Tree (DT)}  is a popular machine learning algorithm for classification and regression tasks that builds a tree-like model of decisions and their possible consequences. The tree consists of nodes that represent decision rules based on the input features, and edges that connect the nodes and represent the possible outcomes or classifications. At each node, the algorithm chooses the input feature that maximizes the information gain, which measures the purity or homogeneity of the data after splitting. Figure 6 shows an example of a classification decision tree adapted from \cite{klaine2017survey}.DT can handle both categorical and continuous input features, and can model non-linear relationships between input features and output variables. DT has several advantages, including interpretability, simplicity, and ability to handle noisy and missing data. However, DT can suffer from overfitting and instability due to the high variance and sensitivity to small changes in the data, which can be mitigated by using techniques such as pruning, ensemble methods (e.g., Random Forest), and regularization.In self-healing scenarios, tree algorithms are basically used to detecting cell outage \cite{omar2021novel}.

5.1)Random forest is a type of ensemble learning method that creates multiple decision trees and aggregates their predictions to produce a final result. The trees in the random forest are constructed using a technique called bagging, where different subsets of the training data are used to train each tree. Additionally, each tree is constructed by selecting a random subset of features to split on, which helps to reduce the correlation between the trees and improve the performance of the algorithm. Random forest is known for its high accuracy, scalability, and ability to handle large datasets with high dimensionality. In the paper \cite{asghar2021artificial}, the authors use this algorithm for detecting cell outage.
\begin{figure}[H]
    \centering
    \includegraphics[width=8cm]{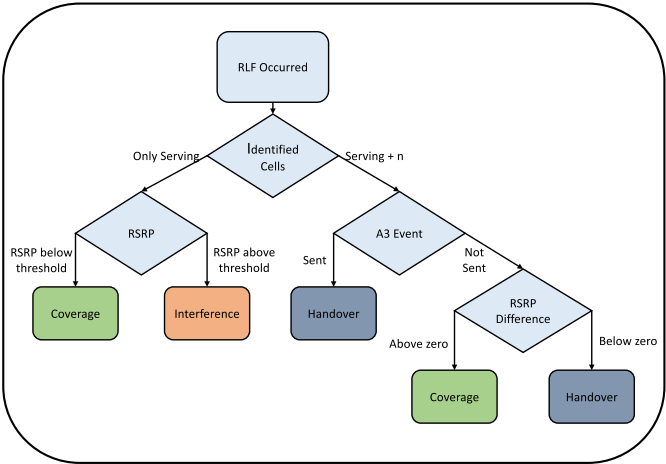}
    \captionsetup{font=small}
    \caption {The figure that is adapted from \cite{klaine2017survey} is an example of a decision tree classification. In this problem, after a Radio Link Failure (RLF) occurred, the algorithm will try to identify the cause of the problem based on other measurements, such as Reference Signal Received Power (RSRP) and Signal to Interference plus Noise Ratio (SINR). Based on these measurements and comparing the RSRP with threshold and measuring its difference, the RLF events are then classified into one of three possible classes.}
    \label{fig:DT}
\end{figure}

\textbf{6)Time-series forcasting}
 Time-series forecasting is a common application of machine learning used to predict future values of a time-dependent variable based on historical data. Time-series data is characterized by having a sequence of data points ordered in time, which makes it a unique type of data that requires specific techniques and models for analysis. Two popular machine learning algorithms used for time-series forecasting are Hidden Markov Models (HMMs) and Prophet Algorithm.

6.1)HMMs are a type of probabilistic model that can be used for time-series forecasting. HMMs model the time-series as a sequence of hidden states and observed variables. The hidden states represent the underlying structure of the time-series, while the observed variables are the actual data points. HMMs have been successfully used for a wide range of applications, such as speech recognition, weather forecasting, and finance. In self-healing, HMM can be found in \cite{alias2016efficient} for detecting cell outage.

6.2)Prophet Algorithm is another popular algorithm for time-series forecasting, developed by Facebook's Core Data Science team\cite{reshmi2021improved}. It is a supervised machine learning algorithm that models the time-series by decomposing it into trend, seasonality, and holiday effects. Prophet uses a Bayesian framework to fit the trend and seasonal components to the data and generate future forecasts based on those components. Prophet is particularly useful for business forecasting, such as predicting sales figures, website traffic, or user engagement metrics. It is known for its ability to handle missing data and outliers in the time-series data, making it robust and flexible. This algorithm is used to detect cell outage in \cite{reshmi2021improved}

\textbf{7)Neural Networks (NN)} are a class of machine learning algorithms inspired by the structure and function of the human brain. They consist of interconnected nodes, or neurons, that work together to learn patterns and relationships in data. There are several types of neural networks, each with their own architecture and strengths.

7.1)Feedforward neural networks (FFNN), also known as multi-layer perceptrons, are the most common type of neural network. They consist of one or more layers of neurons, where each neuron receives inputs from the previous layer and applies an activation function to produce an output. FFNNs are often used for tasks such as image classification, natural language processing, and time series prediction. Currently, the application of neural networks in cell outage management is limited, however an example of the use of a feedforward neural network (FFNN) in detecting cell outages being reported in literature \cite{feng2015cell}. We will later explore in this paper the other related work which have utilized deep neural networks (NNs) in cell fault management.

7.2)Recurrent neural networks (RNN) are a type of neural network designed for sequential data, where the order of inputs matters. Unlike FFNNs, RNNs have loops in their architecture that allow them to maintain an internal state, or memory, which enables them to process sequences of varying lengths. RNNs are commonly used in tasks such as speech recognition, machine translation, and music generation.

7.2.1)LSTM (Long Short-Term Memory) is a type of recurrent neural network architecture that is well-suited for modeling sequential data. It is particularly effective in capturing long-term dependencies and can be used to predict outcomes based on past events. The LSTM architecture includes memory cells that can selectively forget or retain information over time, which allows the network to capture both short-term and long-term patterns in the input data. This makes it a popular choice for various applications, including speech recognition, natural language processing, and time series forecasting. The LSTM architecture has been shown to outperform traditional methods for sequential data analysis and has become a widely used tool in the field of deep learning. In the context of self-healing, article \cite{ouguz2019femtocell} utilizes this model for outage detection.

\textbf{B. Unsupervised Learning (UL): }
This kind of learning can be extremely useful when the
network management function requires identifying anomalous behaviours, recognizing patterns or reducing the dimensionality of the data.
UL is a ML technique where the model is trained on a data set that does not have labeled output or known target variable. The goal of unsupervised learning is to identify patterns, structures, or relationships within the data that are not immediately apparent. This is often achieved through clustering algorithms that group similar data points together or through dimensionality reduction techniques that identify the most important features or variables within the data set. Unsupervised learning is useful for discovering patterns in large and complex data sets, identifying outliers or anomalies, and for feature engineering to prepare data for downstream supervised learning tasks.

\textbf{1) Clustering}.Clustering is a method of unsupervised machine learning that involves grouping together data points based on their similarity. The goal of clustering is to partition a data set into subsets, or clusters, where data points within the same cluster are more similar to each other than to those in other clusters. Clustering algorithms can be hierarchical, where clusters are nested within one another, or flat, where clusters are independent of each other. The most common clustering algorithms include k-means, hierarchical clustering, and DBSCAN. Clustering is widely used in data mining, pattern recognition, image analysis, and market segmentation. It can help identify hidden structures within data, discover unknown patterns, and reduce the dimensionality of data. Clustering can be challenging, especially when dealing with high-dimensional, noisy, or heterogeneous data, and choosing the appropriate algorithm and number of clusters can be subjective and require domain expertise.
\begin{figure}[H]
    \hspace*{0.75cm}
    \includegraphics[width=6.2cm]{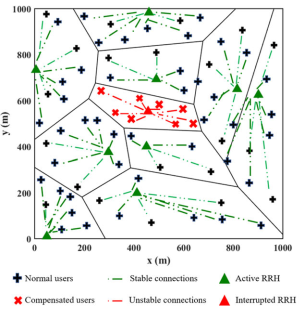}
    \captionsetup{font=small}
    \caption {Initial states when BS outage occurs. Adapted from \cite{yu2020deep}.}
    \label{fig:3}
\end{figure}
1.1)K-means is used to group data points into k distinct clusters based on similarity. It works by iteratively assigning data points to their nearest centroid and moving the centroid to the mean of the assigned data points. K-means is efficient but sensitive to initial placement of centroids and determining the optimal number of clusters. In self-healing, this algorithm is mainly used to detect cell outage as articles \cite{yu2020deep} and \cite{guo2019deep} have utilized k-means for this purpose. The article \cite{yu2020deep} focuses solely on users who are within range of the base station that has been interrupted, as well as its neighboring base stations. 

\begin{figure}[H]
    \centering
    \includegraphics[width=8cm]{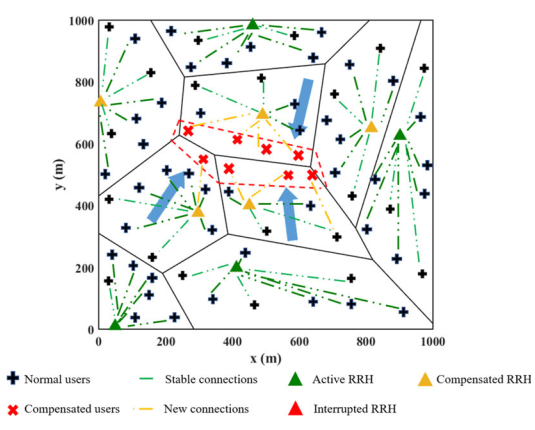}
    \captionsetup{font=small}
    \caption {user reallocation with K-means clustering and DQN algorithm}
    \label{fig:4}
\end{figure}
Based on the illustration in Figure 4, there are 8 compensation users who are being served by the interrupted RRH (Remote Radio Head) (BS). These users have been categorized based on the location of adjacent base stations and are allocated to five adjacent base stations using the K-means clustering algorithm. Figure 5 illustrates the new connections between all base stations and users, and three adjacent base stations have been designated as the compensated base stations for the purposes of their simulation.

1.2)DBSCAN is a density-based clustering algorithm used to group together data points that are closely packed together, while separating those that are further apart. It works by defining a neighborhood around each data point, based on a user-defined distance threshold, and then grouping together dense clusters of neighboring points while separating points in sparser regions. DBSCAN does not require the user to specify the number of clusters beforehand, and is robust to outliers and noise. However, it can struggle with datasets that have varying densities or clusters with non-uniform shapes. In article \cite{ali2018self}, the authors have used this algorithm to detect cell outage.

\textbf{2) Outlier detection}, also known as anomaly detection algorithms, are used to identify data points that are significantly different from the rest of the data. Outliers can be caused by measurement errors, data corruption, or truly rare events that require special attention. Outlier detection algorithms are commonly used in various fields, including finance, healthcare, cybersecurity, and fraud detection. There are many types of outlier detection algorithms, including statistical methods, clustering-based methods, and distance-based methods. 

2.1)Local Outlier Factor (LOF) is one of the popular unsupervised outlier detection algorithms. LOF measures the local density of a data point compared to its neighbors and identifies points that have a significantly lower density as outliers. LOF is widely used in fraud detection, intrusion detection, and industrial quality control, among it application in cell outage detection \cite{onireti2015cell}, \cite{yu2018self}.

\textbf{3)Autoencoders} are a type of Neural Network that can be used for unsupervised learning tasks such as dimensionality reduction, feature extraction, and data compression. The network consists of an encoder that maps the input data to a lower-dimensional latent space representation, and a decoder that reconstructs the original data from the latent space. By minimizing the difference between the input and output data, autoencoders learn to capture the essential features of the input data and can be used for tasks such as anomaly detection.
\begin{figure}[H]
    \centering
    \includegraphics[width=8cm]{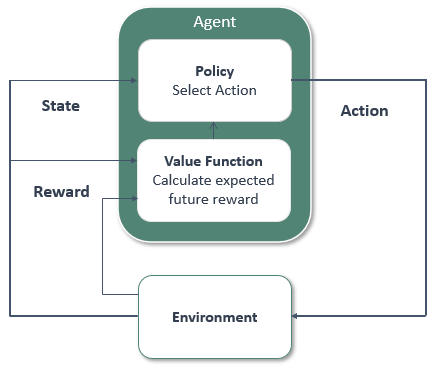}
    \caption{Reinforcement learning architecture}
    \label{fig:5}
\end{figure}
\textbf{\textit{\newline C. Reinforcement Learning (RL):}}
The machine learning methods in this group can be applied to handle network functions that need control over parameters. Unlike supervised learning, the objective of reinforcement learning is to figure out how to accomplish a specific objective through interactions. In numerous practical scenarios, especially those involving sequential decision-making and control, it is not feasible to provide direct guidance to the learning process, such as the correct solution to the problem.
This method is a machine learning approach where an agent interacts with an environment and receives rewards or penalties for its actions, which in turn influence its future actions. The agent's behavior is determined by a policy that balances the benefits of exploiting its current knowledge against exploring new actions that may lead to greater rewards in the future. The agent's expected future rewards are estimated by a value function, which provides an evaluation of the total reward for taking an action in a particular state of the system. In RL, the agent's interaction with the environment is represented as a finite sequence called an episode. Figure 3 depicts the interaction between the agent and environment in RL, with the agent receiving a reward signal for taking an action based on the state of the environment.

In COC, two popular RL techniques are Q-learning and Actor-Critic learning. Both are based on the Temporal Difference approach, where the expected reward is evaluated by looking ahead one step only, and do not require a model of the environment.

\textbf{1) Q-learning} is used to find the optimal action-selection policy based on the maximum expected cumulative reward. The goal of Q-learning is to learn a Q-function, denoted by Q(s, a), which maps each state-action pair to its expected cumulative reward. At each time step, the agent observes the current state, takes an action, observes the resulting reward, and transitions to the next state. The agent updates the Q-function using the following update rule:

$Q(s_t, a_t) \leftarrow Q(s_t, a_t) + \alpha(r_t + \gamma \max_{a} Q(s_{t+1}, a) - Q(s_t, a_t))$

where $s_t$ and $a_t$ are the state and action at time step t, $r_t$ is the reward received after taking action $a_t$ in state $s_t$, $\alpha$ is the learning rate, and $\gamma$ is the discount factor. The update rule adjusts the expected cumulative reward for the current state-action pair based on the observed reward and the maximum expected cumulative reward for the next state. The agent continues to take actions and update the Q-function until it converges to the optimal policy. Q-learning has been used in various applications such as game playing, robotics, and autonomous driving. In the self-healing context, the authors in \cite{adel2021cell}, used this algorithm for COC.

1.2)Fuzzy Q-learning(FQL) is an extension of the traditional Q-learning algorithm that utilizes fuzzy logic to handle uncertainty and imprecision in decision-making. It incorporates the concepts of fuzzy sets and fuzzy rules to model the system's behavior and provide a more flexible approach to learning optimal policies. In fuzzy Q-learning, the Q-values are represented by fuzzy sets, and the action selection is based on fuzzy rules that map the current state to a set of possible actions. This approach enables the algorithm to handle situations where the environment's dynamics are not fully understood or when there is insufficient information available to make precise decisions. Fuzzy Q-learning has been applied to various domains, including robotics, control systems, and game playing, demonstrating its effectiveness in dealing with uncertain and complex environments. The application of this algorithm can be found in \cite{zoha2016learning}.

 \textbf{2) Actor-Critic (AC)} algorithm combines elements of both value-based and policy-based approaches. In AC, the agent maintains both a value function and a policy, and uses these to guide its behavior in the environment. The value function estimates the expected long-term reward for a given state-action pair, and the policy specifies the probability distribution over actions for a given state. The AC algorithm consists of two components: the actor, which learns the policy, and the critic, which learns the value function. The critic updates its estimates of the value function based on the difference between the observed reward and the expected value, using the temporal difference (TD) error:

$$\delta_t = r_t + \gamma V(s_{t+1}) - V(s_t)$$

where $r_t$ is the reward received at time step $t$, $s_t$ and $s_{t+1}$ are the current and next states, $V(s_t)$ is the value function for state $s_t$, and $\gamma$ is a discount factor that determines the importance of future rewards. The actor updates the policy based on the TD error, using a parameter $\alpha$ to control the learning rate:

$$\theta_{t+1} = \theta_t + \alpha \delta_t \nabla_{\theta_t} \log \pi_{\theta_t}(a_t|s_t)$$

where $\theta$ represents the parameters of the policy, $\pi_{\theta}(a_t|s_t)$ is the probability of taking action $a_t$ in state $s_t$, and $\nabla_{\theta_t} \log \pi_{\theta_t}(a_t|s_t)$ is the gradient of the log-probability of the selected action with respect to the policy parameters. By updating the policy and the value function simultaneously, the AC algorithm is able to learn both aspects of the RL problem in a more efficient manner than value-based or policy-based methods alone.

\textbf{\textit{\newline D. Deep Learning (DL):}}

Classical machine learning algorithms have limitations in their ability to solve complex problems with high accuracy. These algorithms often rely on manual feature engineering, which can be time-consuming and may not capture all relevant information in the data. Additionally, these algorithms may struggle to recognize patterns in data with high variability or noise. Deep learning or deep neural network (DNN), on the other hand, can automatically extract features from raw data, allowing it to learn and generalize from vast amounts of information. This makes deep learning particularly useful for applications such as image and speech recognition, natural language processing, and predictive analytics. However, deep learning models can be computationally expensive and require large amounts of data to train effectively. Moreover, the lack of interpretability in deep learning models remains a challenge, making it difficult to understand how these models make predictions. Despite these limitations, deep learning continues to show great promise in advancing the field of artificial intelligence.

Given the importance of self-healing in modern communication networks, deep reinforcement learning has emerged as a promising approach for effectively managing network resources and responding to changing network conditions. In particular, the application of deep reinforcement learning to cell outage compensation, radio resource management, and dynamic spectrum sharing has shown significant promise for improving the efficiency and reliability of 5G and 6G networks. Therefore, in the context of self-healing, it is essential to focus on the development and implementation of deep reinforcement learning algorithms that can effectively adapt to changing network conditions and improve the overall performance and resilience of communication networks. 

On the other hand, one limitation of RL techniques is that the state-action table size can become a scalability issue, as it is proportional to the number of system states and possible actions. To overcome this limitation, deep RL techniques utilize a deep neural network to perform the mapping from states to actions that is illustrated in figure 7. The neural network is trained using the basic RL technique to determine the action with the highest reward for each state, based on the RL subsystem's experience. To reduce the effect of specific conditions during an episode, the state-action-reward data for each episode is stored in a replay memory. This allows the neural network to be randomly trained using samples from multiple episodes.
\begin{figure}[H]
    \centering
    \includegraphics[width=8cm]{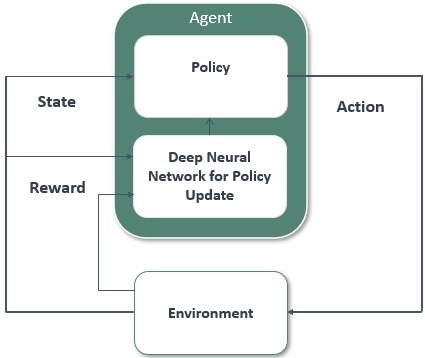}
    \caption{DRL with neural network for policy update}
    \label{fig:6}
\end{figure}
The examples of deep reinforcement learning algorithms that will be discussed demonstrate the potential of this approach in addressing critical challenges in self-healing 5G and 6G networks.

\textbf{1) Deep Q-Network (DQN)} is a DRL algorithm that combines deep neural networks with Q-learning. DQN has been used to solve various problems, including game playing, robotics, and natural language processing. The algorithm works by using a deep neural network to estimate the Q-values of each possible action given the current state. The network is trained using a variant of Q-learning called experience replay, which involves storing the agent's experience in a replay memory and sampling a batch of experiences randomly to update the network's parameters. DQN is known for its ability to learn directly from raw sensory inputs, making it useful for problems where hand-crafted features are not available or are difficult to design. In the \cite{yu2020deep} the author has applied DQN to solve COC problem.

We discuss later on in this
paper how ML techniques have been used in cell outage management.

\begin{table*}
\centering
\caption{SUMMARY OF SELF-HEALING USE CASES IN TERMS OF ML METHODS}

\begin{center}
\begin{tabular}{ |l|l|l|l|l| }
\hline
\multicolumn{5}{ |c| }{Self-Healing} \\
\hline
 & Ref+Y & ML technique &  Objective &  Algorithm \\ \hline
 \multirow{5}{*}{COD} & \cite{onireti2015cell} 2015 & SL/UL & Anomaly detection & k-NN, LOF  \\
 &\cite{feng2015cell} 2015 & SL & Diagnosis & FFNN \\
& \cite{fortes2015contextualized} 2015 & SL & Diagnosis &  Naive Bayesian \\
 & \cite{alias2016efficient} 2016 & UL & Pattern identification & Hidden Markov Model \\
  & \cite{zoha2016learning} 2016 & RL & Anomaly detection & SVM and LOF \\
 & \cite{yu2018self} 2018 & UL & Pattern identification & LOF \\
 & \cite{ali2018self} 2018 & SL/UL & Anomaly detection & DBSCAN algorithm \\
 & \cite{asghar2019assessment} 2019 & UL & Pattern identification & DL, Autoencoder \\
 & \cite{ouguz2019femtocell} 2019 & SL & Pattern identification & LSTM \\
& \cite{reshmi2021improved} 2021 & SL &  Anomaly prediction & Prophet Algorithm \\

 & \cite{omar2021novel} 2021 & SL & Pattern identification & Decision tree \\

 & \cite{asghar2021artificial} 2021 & SL & Pattern identification & DT, SVM, Random Forest \\
 & \cite{ruan2021intelligent} 2021 & SL & Pattern identification & SVM, NN \\
& \cite{yu2022active} 2022 & SL & Pattern identification & Hierarchical clustering \\

 \hline
\multirow{4}{*}{COC} & \cite{moysen2014reinforcement} 2014 & RL & Control Optimization & Actor-Critic \\
 & \cite{onireti2015cell} 2015 & RL & Control Optimization & Actor-Critic \\
& \cite{zoha2016learning} 2016 & RL & Coverage Optimization & FQL \\

 & \cite{yang2019deep} 2019 & DRL & Maximizing the throughput & K-means and DQN \\
 & \cite{yu2020deep}/\cite{guo2019deep} 2020/2019 & DRL & Control Optimization & K-means and DQN \\
 & \cite{adel2021cell}2021 & RL & Control Optimization & Q-Learning \\
 & \cite{vaezpour2022deep}2022 & SL & Control Optimization & DNN \\

\hline
\end{tabular}
\end{center}
\end{table*}

\section{\Roman{section}. Self-healing solutions applied in cellular networks}
This section will discuss the related works that have successfully applied ML methods in self-healing, focusing on cell outage compensation (COC) and cell outage detection (COD) that have been presented in Table 1.

\begin{figure}[H]
    \centering
    \includegraphics[width=8cm]{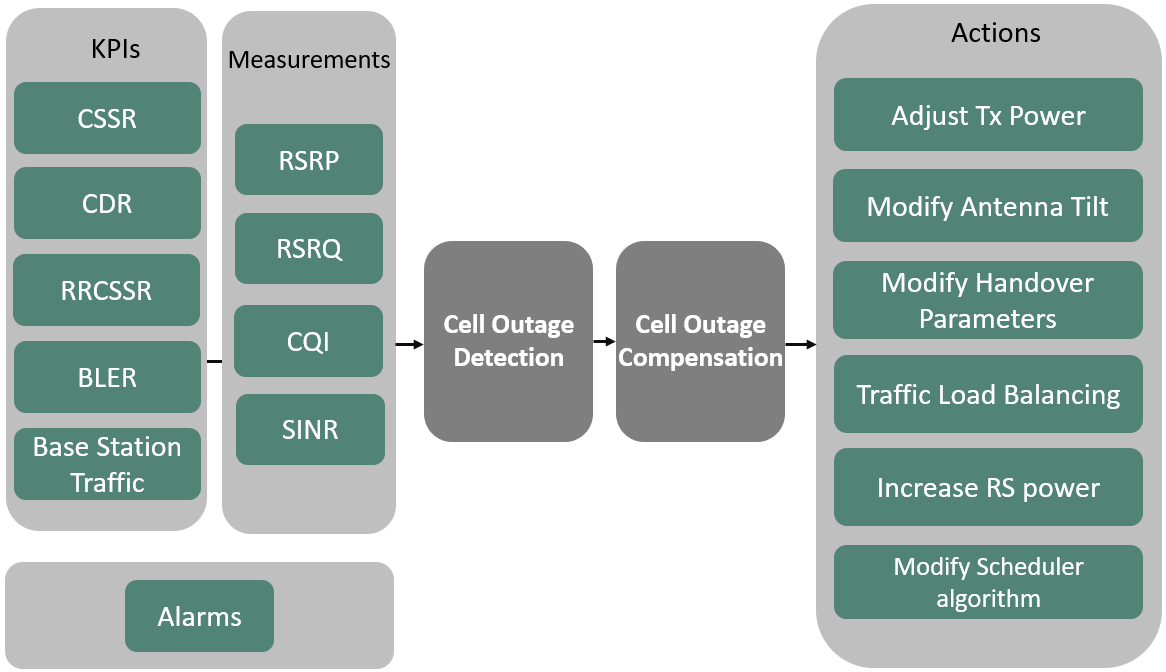}
    \caption{Network inputs and action space for Self-Healing}
    \label{fig:9}
\end{figure}

\textbf{\textit{Cell Outage Detection:}}
As mentioned before, the main goal of cell outage detection is to identify potentially malfunctioning cells. To achieve this, various current studies employ Key Performance Indicator (KPI) parameters including the number of handover requests received by the base station, Call Setup Success Rate (CSSR), Call Drop Rate (CDR), Radio Resource Control Connection Setup Success Rate (RRCSSR), Block Error Rate (BLER) and  Base station traffic. These KPIs and Measurement parameters such as Reference Signal Received Power (RSRP), Reference Signal Received Quality (RSRQ), Channel Quality Indicator (CQI), Signal to Interference and Noise Ratio (SINR) are used to detect and diagnose cell outages. In figure 8 these parameters have been mentioned.
 
Machine learning techniques relevant to 5G self-healing and the progress made in applying them are discussed in \cite{zoha2015data}. In \cite{zoha2016learning}, an optimized decision tree was investigated when using a Quality of Experience (QoE) key performance indicator to diagnose dysfunctional network nodes. While, \cite{fourati2021comprehensive} discusses the architectural concepts for implementing a decision tree for fault detection.
Anomaly detection algorithms, moreover, provide a compelling solution to this issue, as they can pinpoint unusual measurements that may indicate an underlying issue within the network. Consequently, various researchers have explored this particular use case, such as in \cite{onireti2015cell}, Onireti et al. used a KNN and local-outlier-factor-based anomaly detection for the control plane and a heuristic grey prediction approach for anomaly detection for the data plane. The experiments have shown that it can reliably detect both control and data plane outages.

Moreover, the paper \cite{zoha2016learning} analyzes two different anomaly detection algorithms in order to detect the outage, LOF and SVM. 

Additionally, in \cite{ali2018self} the authors introduced the wider context of resilience in 5G networks and presented an anomaly detection and diagnosis based self-healing concept based on DBSCAN algorithm, which enables early detection of and reaction to problems in a dynamic way. 

To identify an outage, the authors of \cite{asghar2021artificial}, compare and analyze the performance of random forest, SVM, and KNN, in order to identify anomalies in the reports of user measurements.
In order to determine if there is a failure. 

The paper \cite{reshmi2021improved} proposes an automated network diagnostic and self-healing technique for 5G environments using predictive analysis. The system collects data on device or network performance parameters and analyzes for possible anomalies. Predictive analysis is then used to diagnose problems in the network when performance parameters deviate from normal ranges. Time series analysis is used to predict network performance in various time intervals. The proposed technique was implemented in a live network environment provided by a leading ISP, and the performance analysis showed that the predictive analysis and network diagnostics improved network performance with self-healing in 5G networks. The paper introduces the use of the Prophet algorithm for time series analysis.

In reference \cite{asghar2019assessment}, a solution for detecting cell outages was presented utilizing a Neural Network, specifically an autoencoder. The method employed simulated data generated from a SON simulator, and its performance was evaluated by comparing it to the results obtained from the k-NN ML technique. A similar use of Neural Network can be found in \cite{ruan2021intelligent}, where the authors use RRCSSR, CDR and BLER parameters to complete the
function of outage detection.

In the study described in \cite{alias2016efficient}, a solution for detecting cell outages was created using a Hidden Markov Model. This approach involves observing and capturing the current states of various base stations (BSs) to estimate the occurrence of a Cell Outage. 
The study \cite{ouguz2019femtocell} uses metrics such as Signal to Interference Noise Ratio (SINR) and Channel Quality Indicator (CQI) as input features for the LSTM structure to detect outage patterns based on time sequences of UE data around femtocell sites. The LSTM network is trained on UE data and tested on test data, achieving a 77\% accuracy in predicting femtocell outage states.
Article \cite{yu2018self} proposed a solution for detecting cell outages in 5G H-CRAN, utilizing a modified version of the Local Outlier Factor unsupervised anomaly detection algorithm. This approach is more efficient than the traditional Local Outlier Factor algorithm. Additionally, the proposed solution has a low false negative rate. The study also presented a centralized self-organized COD architecture that incorporates the modified Local Outlier Factor algorithm to improve accuracy and intelligence, even in the presence of multiple errors in cell outage consideration. 
From the initial studies, the Bayesian Networks/Naive Bayesian Classifier method has also been a commonly employed approach for diagnosis. This method typically requires an expert to define the logical relationships that form the network, but the necessary probabilities can then be extracted from historical data if available, as seen in \cite{fortes2015contextualized}. The Naive Bayesian Classifier method assumes that only one cause is present at a time, and that the symptoms are independent given the cause. While this assumption may not always be realistic for certain faults in an actual network, studies have reported acceptable diagnostic performance for the scenarios examined.
In a recent study \cite{yu2022active} authors have proposed a clustering method that can group users in a cell based on their performance, providing a way to determine the state of the base station. The method uses a hierarchical clustering algorithm, specifically a bottom-up approach, which eliminates the need for selecting an initial point or cluster number. The algorithm treats each user as a cluster and successively merges similar clusters until a termination condition is met, effectively building a cell outage detection mechanism for broadband services in 5G C-RAN.

\textbf{\textit{Cell Outage Compensation:}}
The objective of compensation, as mentioned earlier, is to recover the highest achievable level of service utilizing the available network resources while following the priorities defined by the network operator's policy. The majority of the efforts have been focused on compensating for cell outages. In this situation, the most widely used method is to adjust Tx power, the antenna tilt settings of the adjacent base stations, modify Handover parameters, traffic load balancing, increase RS power,  and modify the scheduler algorithm. In figure 8 the set of most used actions has been mentioned. For this use case RL has been proved as a valid solution since it is a continuous decision making/control problem (see figure 9).

In this specific context, the authors of \cite{onireti2015cell} and \cite{moysen2014reinforcement} have presented a significant contribution to the field of self-healing. They have proposed a comprehensive solution that automatically mitigates the impact of outages by adjusting relevant radio parameters of nearby cells. Their approach involves optimizing coverage and capacity in the affected outage zone by modifying the electrical tilt of the antenna and the downlink transmission power of the adjacent eNBs. To implement their proposed method, the authors suggest using a RL algorithm based on actor-critic theory, which can make online decisions at each eNB and adapt to changes in the situation such as user mobility, shadowing, and the decisions made by surrounding nodes to address the same problem.
While the paper\cite{adel2021cell}, proposes a novel Q-learning-based algorithm to compensate for the coverage gap in the outage area by modifying the power and antenna tilt angle parameters of neighboring cells. Unlike existing COC approaches, the proposed algorithm does not assume knowledge of the mathematical models of the system and guarantees fully autonomous and accurate COC by learning the consequences of the taken actions to compensate for the coverage gap. This makes the proposed algorithm highly adaptive and suitable for different environments. The paper's contributions include introducing an autonomous and accurate COC method and demonstrating the effectiveness of Q-learning in SONs.
\begin{figure}[H]
    \centering
    \includegraphics[width=8cm]{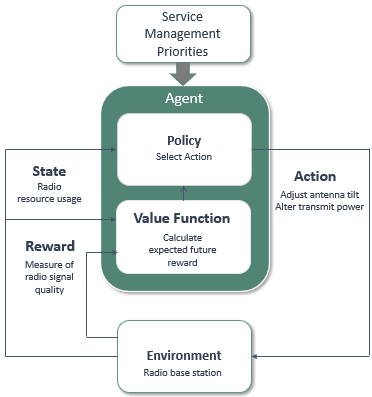}
    \caption{Compensation using reinforcement learning}
    \label{fig:7}
\end{figure}

Zoha et al. \cite{zoha2016learning} presented a study utilizing Fuzzy Q-learning (FQL) to address cell outage compensation. The authors introduced a framework that relies on MDT (minimise drive testing) functionality measurement reports to detect outages and mitigate their impact. The MDT measurements are initially collected and reduced in dimension using MDS (Multi dimensional Scaling) for outage detection. To compensate for the outage, the authors propose a mechanism that combines a Fuzzy controller with RL to adjust the antenna down-tilts and transmission powers of surrounding cells, minimizing the impact of the outaged cell.

In \cite{yu2020deep}, \cite{guo2019deep} authors apply K-means clustering to allocate compensation users to nearby base stations and DQN to determine the optimal antenna downtilt and power allocation. The simulation results demonstrate that the proposed algorithm converges rapidly and is stable, reaching 95\% of the maximum target value, and effectively handles cell outage compensation in 5G network.

  A COC contribution also based on DQN is targeted in \cite{yang2019deep}, where the authors propose a framework to address the cell outage scenario in the Ultra-Dense Network (UDN). The proposed framework aims to maximize the sum of the throughput of all users while satisfying the service quality requirements of each mobile user. The framework utilizes the K-means clustering algorithm to allocate compensation users to nearby base stations and employs a deep neural network (DNN) to approximate the action-value function. The simulation results indicate that the proposed algorithm converges quickly and is stable, reaching 99.53\% of the maximum throughput. The paper concludes that the DRL-based framework is efficient and effective in meeting user rate requirements and handling cell interrupt compensation in the UDN.

 Finaly, paper \cite{vaezpour2022deep} proposes a newly NOMA-based (Non-orthogonal multiple access) cell outage compensation scheme to enable a network to react to a cell failure quickly and serve users in the outage zone uninterruptedly. The compensation is formulated as a mixed integer non-linear program (MINLP), where outage zone users are associated with neighboring cells and their power is allocated to maximize spectral efficiency while maintaining the quality of service (QoS) for the rest of the users. To handle the computational complexity, the paper develops a low-complexity suboptimal solution using a newly heuristic algorithm to determine the user association scheme and applying an innovative deep neural network (DNN) to set the power allocation.

 \section{\Roman{section}. Conclusion}
 Many studies have demonstrated the potential of ML techniques in providing automated self-healing functions for fault detection and compensation. The detection of cell outages and subsequent compensation strategies have been the focus of much research, leading to a consensus on the most appropriate ML techniques for these tasks. As a result, we can expect to see rapid advancements in the application of ML techniques for cellular fault management in the near future. In particular, enhanced deep learning approaches that can interact with humans productively and build trust in cases where human involvement is necessary are likely to be a key area of focus. Overall, the promising results seen in the studies reviewed in this paper suggest that ML-based self-healing has the potential to significantly enhance the efficiency and reliability of cellular networks, ultimately benefiting both operators and end-users.

\bibliographystyle{IEEEtran}
\typeout{}
\bibliography{citation.bib}
\end{multicols}
\end{document}